\documentclass[10pt,a4paper]{article}
\usepackage{authblk}

\usepackage{amsmath}
\usepackage{graphicx}
\begin{document}

\title{Simulated hysteresis loop of a bent Fe-rich glass-covered wire}

\author[1]{Przemys{\l}aw Gawro\'nski}
\affil[1]{Faculty of Physics and Applied Computer Science, AGH University of Science and Technology, Cracow, Poland}
\author[2]{Alexander Chizhik}
\affil[2]{Departamento Fisica de Materiales, Facultad de Quimica, Universidad del Pais Vasco, San Sebastian, Spain}
\author[1]{Krzysztof Ku{\l}akowski}

\maketitle   

\abstract{We present new micromagnetic simulations, where an influence of the wire shape on the spatial distribution of internal magnetoelastic stress is taken into account. Local magnetoelastic anisotropy is approximated to be uniaxial, with spatial dependence due to the stress distribution. The hysteresis loop is calculated with the $mumax^3$ }{software for the wire volume and separately for selected parts of the wire, including the surface. The bistability of the calculated surface hysteresis loops depends on the behaviour of the neighboring part of the inner core, most close to the surface element where the surface loop is measured. }

\section{Introduction}
Amorphous micro- and nanoscopic soft magnetic wires have been found applications in sensors and data storage \cite{a1,a2,a3,a4,a5}. Magnetic properties of these wires are known to depend on the spatial distribution of internal and applied stresses \cite{b1,b2}. As a rule, the stresses applied are either tensile or torsional; these kinds of stresses come also from the glass covers and from the fabrication process, as it has been discussed by several authors \cite{c1,c2}.  On the contrary, the role of bending stresses has been largely ignored, with \cite{mv1,mv2,mv3,mv4} as rare exceptions. Yet, bending is known to strongly modify internal distribution of stresses \cite{mv1}, and therefore the radius of curvature of a wire is a convenient parameter to control its magnetic properties.

\paragraph{} Our aim here is to investigate the volume and surface hysteresis loops of a wire in the presence of the bending stress by means of computer simulations. Our motivation is twofold. First is that, as suggested in \cite{vel}, the spatial distribution of stress can depend on the fabrication process. Therefore, stress dependent magnetic properties can vary from one series of samples to another. The method of simulation offers a test-bed, where correlations can be investigated between theoretical stress distribution and the simulated magnetic properties. Second motivation is to control the volume and surface hysteresis loops in the same magnetization process. Such results are of interest when comparing and interpreting experimental data, obtained with the conventional induction method and with the Kerr effect.

\paragraph{} The structure of the text is as follows. In the next section we recall the theoretical and experimental achievements on the magnetism of bent wires, known in literature. In Section 3 we describe the simulation, including an approximation of an effective uniaxial anisotropy, introduced to meet the demands of the used software. There we describe also the parametrization of the spatial stress distribution, based on Refs. \cite{mv1,sto}. Last two sections contain numerical results and their discussion, respectively.

\section{Literature data} In Ref. \cite{mv1}, the conventional hysteresis loop was measured of a Fe-based wire with 125 $\mu m$ diameter, as dependent on the radius of curvature $R_c$; the latter was tuned from 12 cm to about 1 cm. The bistable character of the loops was found to disappear for $R_c$ less than 11 cm, but it was restored for $R_c$ less than about 2.5 cm. The paper provides also theoretical evaluations of the stress distribution within the bent wire, based on the classical theory of elasticity \cite{tim}. 

\paragraph{} In Ref. \cite{mv2}, the hysteresis loops have been measured by the Kerr method for the wires of the same diameter and composition as in Ref \cite{mv1}. The experimental results have been accompanied by theoretical considerations on the energy barrier which is exceeded during the remagnetization process. In the discussion, the Authors highlight the role of closure domains and of the intrinsic stresses frozen-in during the fabrication process. 

\paragraph{} In Ref. \cite{mv3}, the role of wire length is investigated experimentally for the same Fe-based microwires. The samples are wound on cylindrical holders, then the radius of curvature was equal to the radius of the holder. The Authors argue, that the bistability is present for any value of the radius of curvature, if only the wire is long enough. The study deals also with an influence of temperature and the amplitude of the applied field on the character of the loops.

\paragraph{} On the contrary to the literature cited above, Ref. \cite{mv4} concerns a direct application of glass-covered Fe-rich microwires as bending sensors. The wires diameter here varies from 1 to 40 $\mu m$. Instead of the switching (coercive) field, the switching time is measured, what largely improves the sensor sensitivity. It appears that the switching time varies as the square root of bending. The paper contains also an updated list of applications of glass-coated microwires as sensors. More generally, glass-covered micro- and nanoscopic wires offer reduced dimensions, larger field sensitivity and more controllable fabrication process \cite{a5}.

\section{Simulation of a Fe-rich glass-covered wire} 

The Landau-Lifshitz equation is solved numerically by means of the micromagnetic simulation software $mumax^3$ \cite{mum}. The diameter of the discretization grains is 50 $nm$, while the wire diameter is 2 $\mu m$, and the length is 100 $\mu m$. The Landau-Lifshitz damping constant $\alpha$= 0.1, the exchange stiffness constant for iron $A_{ex}=2\times 10^{-11}$ J/m, the magnetostriction constant for the Fe-rich wire $\lambda= 2.5 \times 10^{-6}$, and the saturation magnetization $\mu _0M$ = 0.6 T \cite{a5}. For the bent wire, the radius of curvature $R_c$ is 2 mm. The simulation is quasistatic, which is equivalent to a very low frequency of the field variation.

\begin{figure}[t]%
\centering
\includegraphics[width=.49\textwidth]{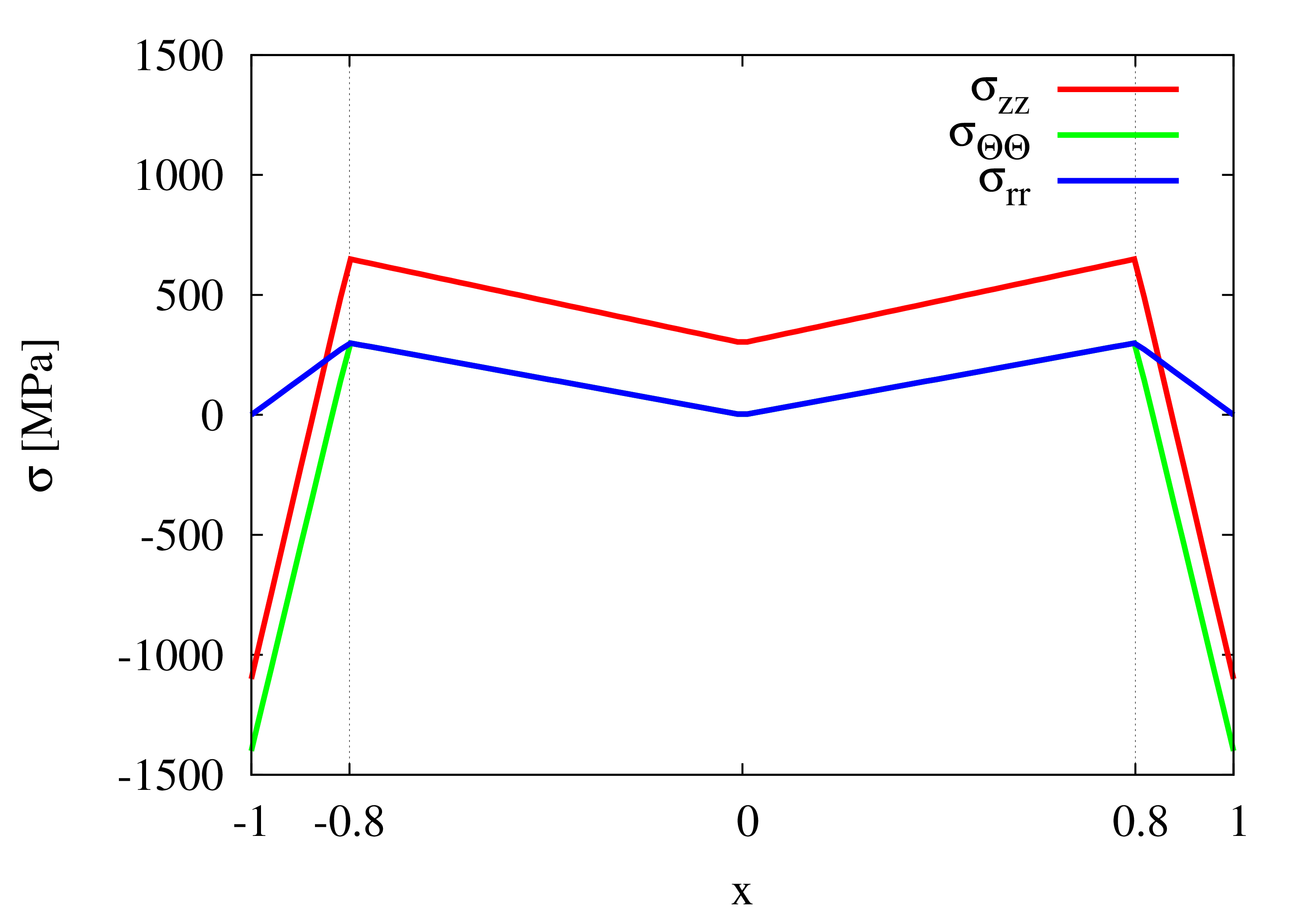}
\caption{The stress distribution in a straight wire. In the horizontal axis, we mark the position on a diameter. The wire radius $r=1 \mu m$.}
\label{fig1a}
\end{figure}

\begin{figure}[t]%
\centering
\includegraphics[width=.4\textwidth]{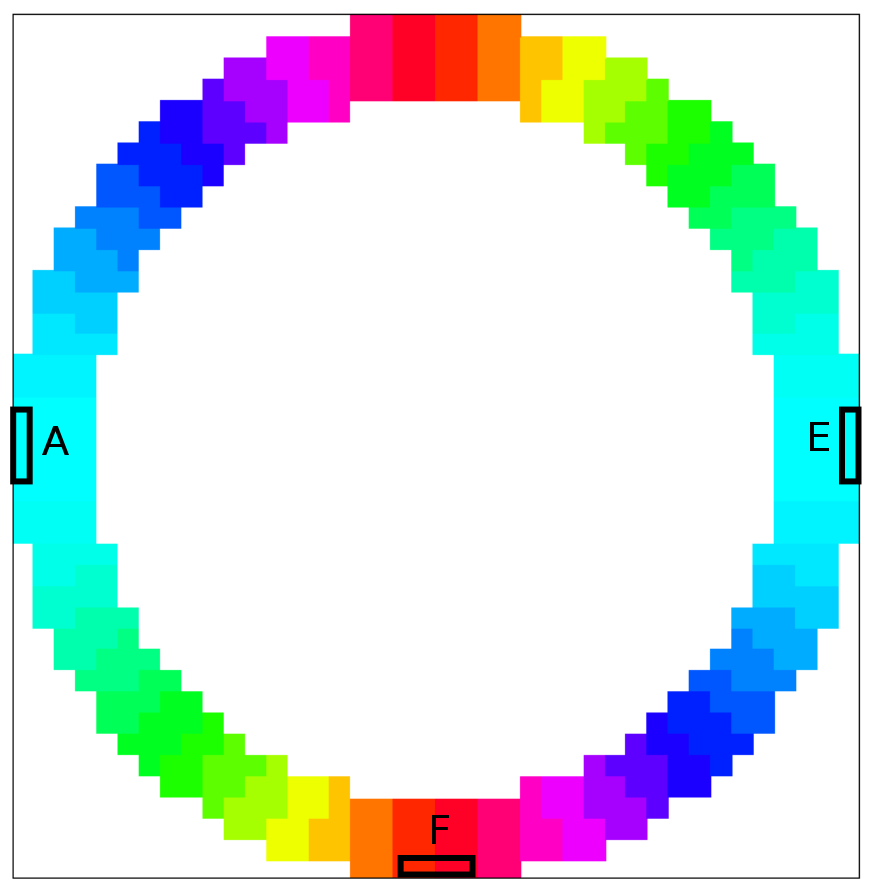}
\caption{The stress distribution in a straight wire, the same as in Fig. \ref{fig1a} - a cross section. White area marks the easy anisotropy axis along the wire. Colors online differentiate radial directions of local anisotropy axes. The positions A, E and F mark the wire fragments, where the hysteresis loops are calculated. These loops are shown in Fig. \ref{surfstr}. }
\label{fig1ac}
\end{figure}

\paragraph{} The software $mumax^3$ limits the forms of magnetic anisotropy to two: uniaxial and cubic one. Then, numerical calculations can be performed only for these two forms. To fulfill this condition, we approximate the local magnetoelastic energy as uniaxial one. The idea is as follows. For a Fe-rich magnetostrictive wire, the cylindrical symmetry allows to limit the number of stress components to three: radial, circumferential and axial one \cite{tdl,sto}. Locally, these three directions can be related to Cartesian coordinates, with three different stresses along each axis. Let us recall that the energy of the magnetic anisotropy is equivalent to a barrier of energy, which is necessary to rotate a local magnetic moment. Its uniaxial character is equivalent to the cylindrical symmetry of the angular dependence of the magnetic moment; if the local easy axis is along OZ direction, the same energy is needed to rotate the magnetic moment within ZX plane and ZY plane. The idea of an effective uniaxial 
anisotropy is that the magnetic moment is rotated within the plane where the energy cost is lower. To be more detailed, suppose that there are three tensile stresses $\sigma_x$, $\sigma_y$ and $\sigma_z$ at a given point of a magnetic sample with positive magnetostriction $\lambda$. Without limiting the generality we can assume that $\sigma_x$ $<$ $\sigma_y$ $<$ $\sigma_z$; hence the easy magnetic axis is along OZ. The idea is to select the uniaxial anisotropy constant as $\lambda (\sigma_z-\sigma_y)$. This choice is determined by the fact that the energy to rotate the magnetic moment in the plane ZY is less than the one to rotate it in the plane ZX.

\begin{figure}[h]
\centering
\includegraphics[width=.49\textwidth]{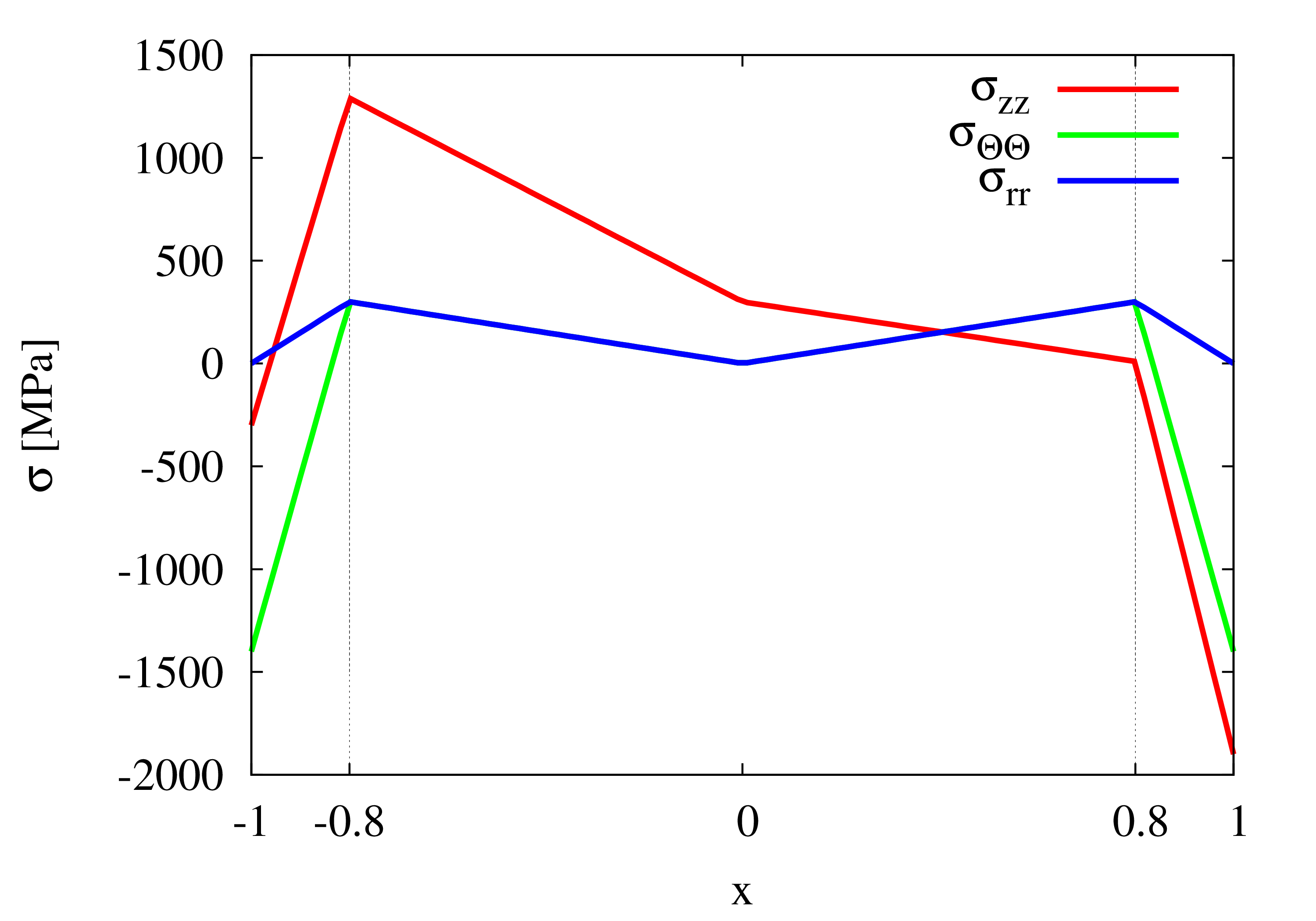}
\caption{The stress distribution in a bent wire. In the horizontal axis, we mark the position on a diameter. The wire radius $r=1 \mu m$, and the radius of curvature $R_c$=2 mm.}
\label{fig1b}
\end{figure}

\begin{figure}[t]
\centering
\includegraphics[width=.4\textwidth]{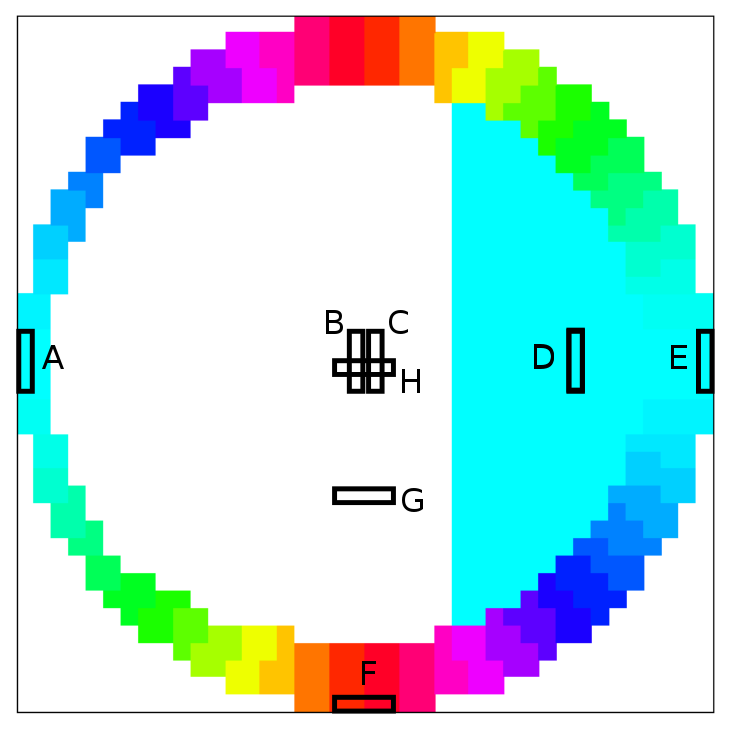}
\caption{The stress distribution in a bent wire -- a cross section. The distribution is as in Fig. \ref{fig1b}, with the orientation of the stress along OX (for details see the text). White area marks the easy anisotropy axis along the wire. The positions marked with labels indicate the wire fragments, where the hysteresis loops are calculated, as shown in Figs. \ref{lefrig} and \ref{downup}. The color online on the right part of the figure is related to the assumed direction (but not the value) of the magnetic anisotropy in the plane perpendicular to the wire. }
\label{map2}
\end{figure}

\begin{figure}[t]
\centering
\includegraphics[width=.4\textwidth]{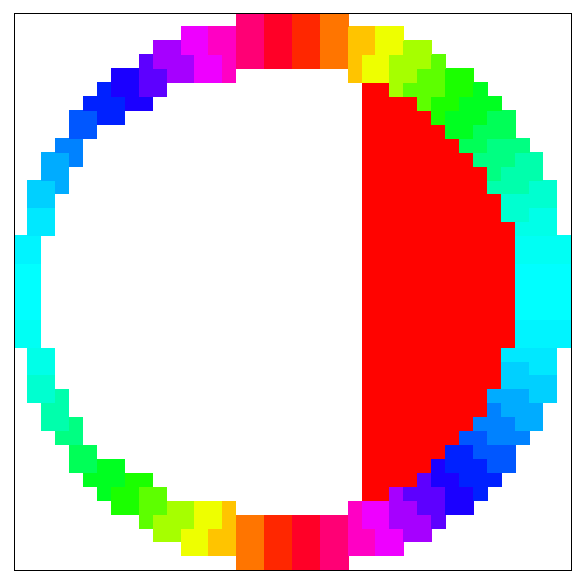}
\caption{The stress distribution in a bent wire -- a cross section. The distribution is as in Fig. \ref{fig1b}, with the orientation of the stress along OY (for details see the text). White area marks the easy anisotropy axis along the wire. The color online on the right part of the figure is related to the assumed direction (but not the value) of the magnetic anisotropy in the plane perpendicular to the wire. }
\label{fig2c}
\end{figure}

\paragraph{} In amorphous wires, the shape magnetostriction coefficient is a scalar \cite{tdl}, and the spatial distribution of the magnetic anisotropy is determined by the distribution of stress. For the case of a straight wire, we parametrize the stress distribution as in \cite{sto}, with small simplifications in order to reduce the number of different discretization grains; the latter is limited by the software $mumax^3$ to 256. The stress distribution is shown in Figs. \ref{fig1a}, \ref{fig1ac}. The same information for the bent wire is shown in Figs. \ref{fig1b}, \ref{map2}, \ref{fig2c}. These pictures are based on the calculations performed in \cite{mv1,tim}. In particular, the most relevant component of the bending stress $\sigma _{zz}$ was found to vary linearly with the distance from the curvature center. Namely, we approximate this dependence as 

\begin{equation}
 \sigma _{zz}=-\frac{2Erx}{R_c}
 \label{sigtet}
\end{equation}
where $E = 1.6\times 10^{5}$ MPa is the Young modulus, $r$ is the wire radius, $R_c$ is the radius of the wire curvature, and $x$ varies within the wire volume from $-1$ at the surface of the convex part to $+1$ at the surface at the concave part. The approximation in Eq. \ref{sigtet} is justified as long as the wire radius $r$ is much less than the radius of curvature $R_c$.  For $r/R_c=5\times 10^{-4}$, the bending component of the stress changes from -800 to +800 MPa.
\begin{figure}[th]%
\centering
\includegraphics[width=.4\textwidth]{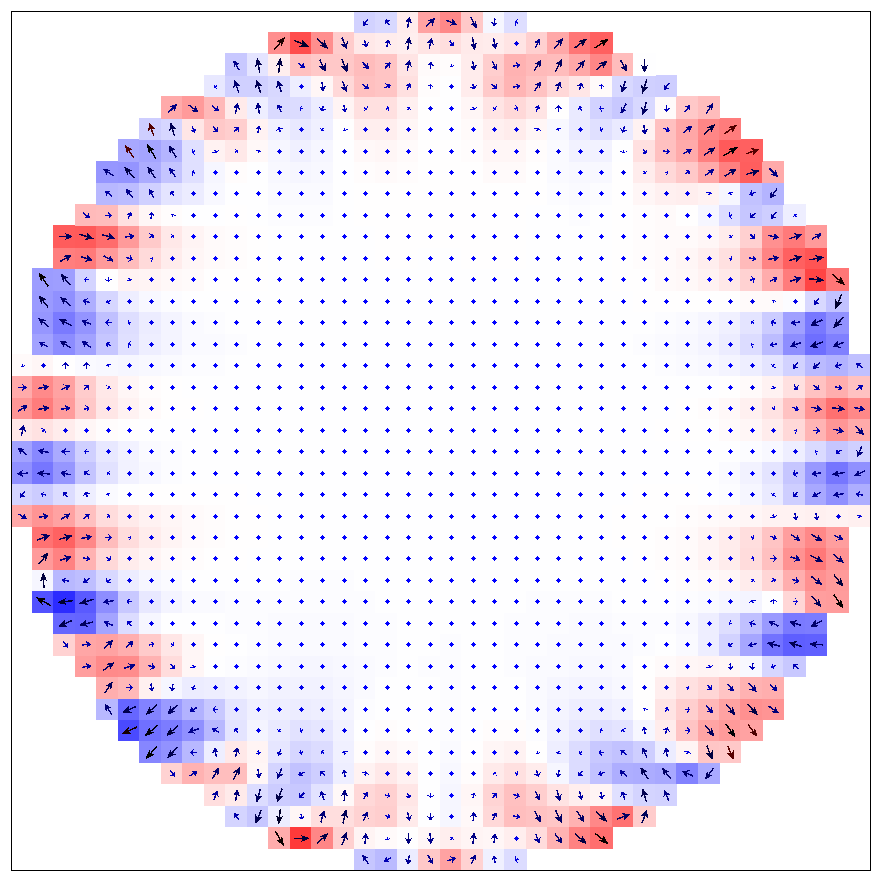}
\caption{%
The magnetization distribution in a cross section of a straight wire, at the half of the wire length. The colors (red and blue online) mark the horizontal component of the orientation to right and left, respectively. }
\label{dis}
\end{figure} 

\begin{figure}[bph!]%
\centering
\includegraphics[width=.4\textwidth]{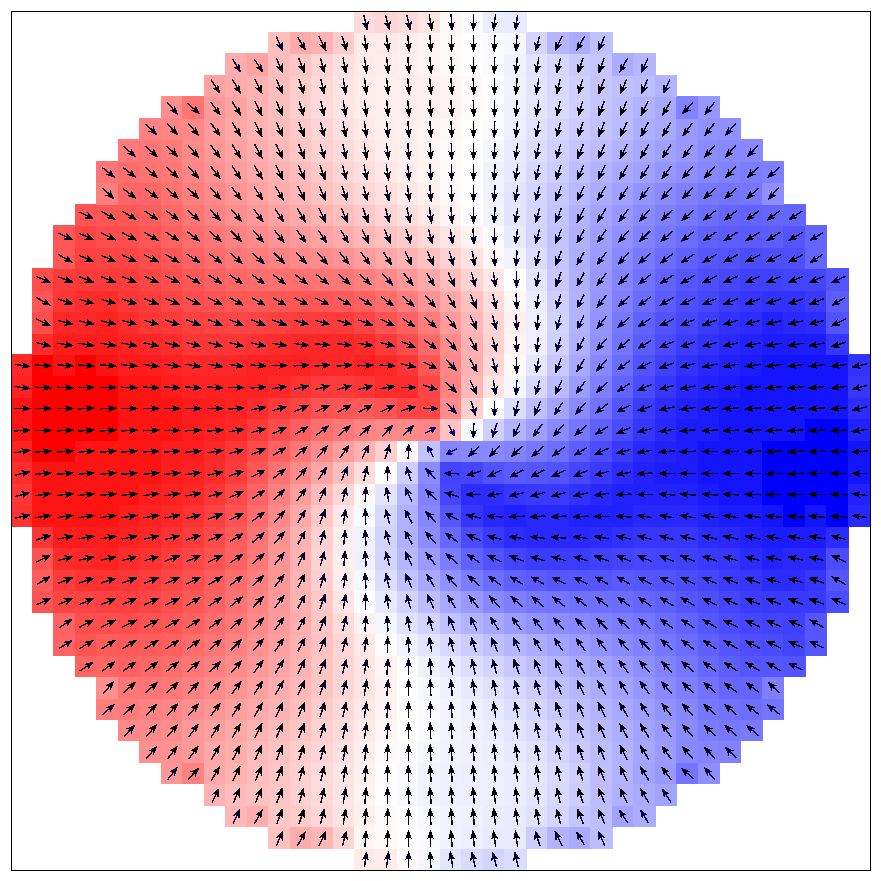}
\includegraphics[width=.4\textwidth]{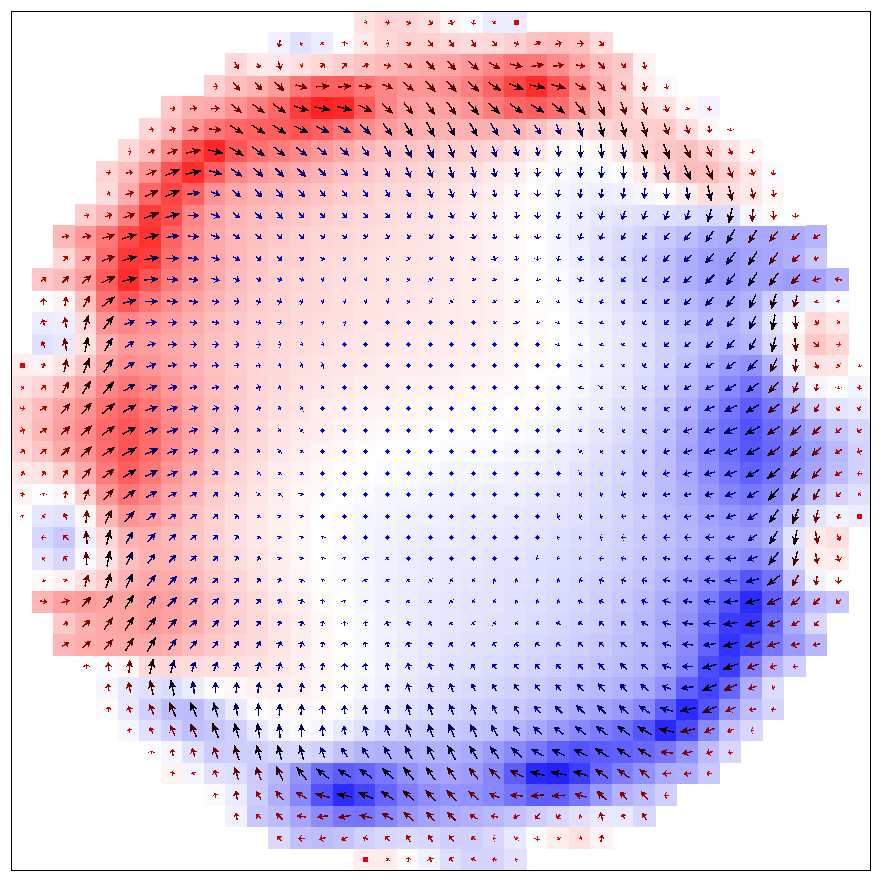}
\includegraphics[width=.4\textwidth]{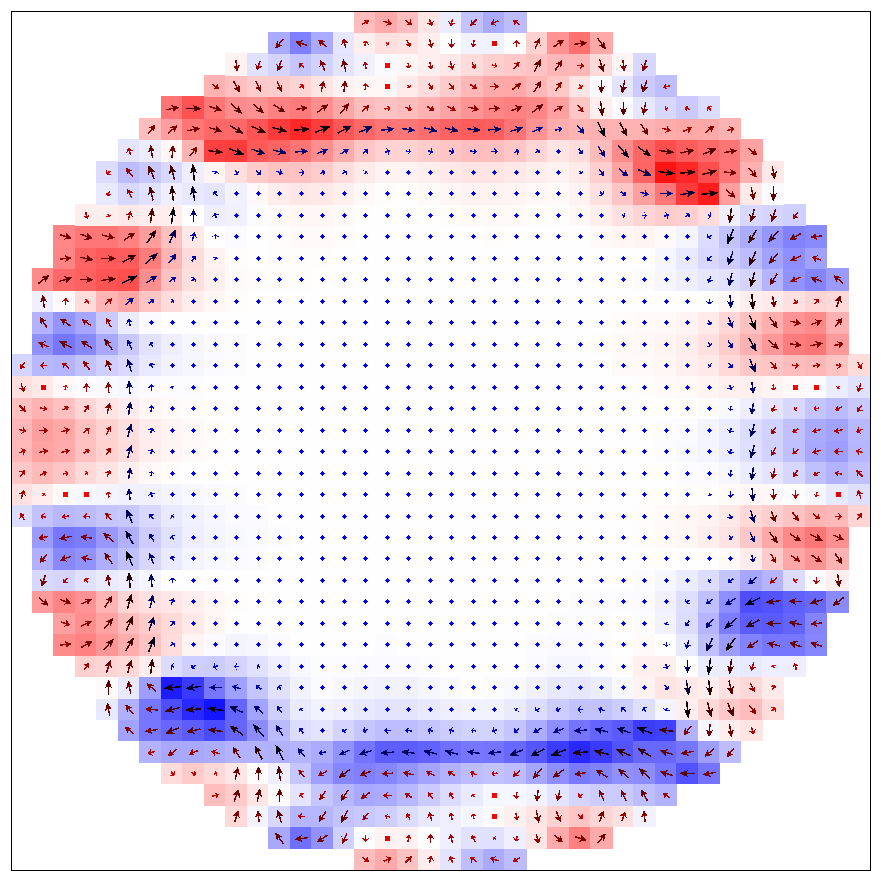}
\caption{The magnetization distribution in a cross section of a straight wire at the distances zero, 0.25 and 0.5 $\mu m$ from the wire end from top to bottom. The colors (red and blue online) mark the horizontal component of the orientation to right and left, respectively. }
\label{dis2}
\end{figure} 

\paragraph{} Basically, we expect that in the compressed part of the wire a competition exists between the stresses $\sigma_{\theta \theta}$ and $\sigma_{rr}$. This competition leads to a different direction of effective stress at each point of the wire. This fact is unfavourable, as we have to keep in computer memory each grain of discretization. To evade this difficulty, we assume a homogeneous direction of the magnetic anisotropy (but not the value) in the compressed part of the inner core. These two versions of the model distribution of stress are shown in Figs. \ref{map2}, \ref{fig2c}. To check this assumption, calculations on the bent wire are performed twice: for this direction either OX (Fig. \ref{map2}) or OY (Fig. \ref{fig2c}). Below we show that the results do not differ remarkably; hence the assumption is justified. 

\section{Results}  In Fig. \ref{dis} we show the spatial distribution of the magnetization in a cross section of a straight wire of radius $r = 1 \mu m$, at half of its length $L$ ($L = 100 \mu m$). Both the outer shell, with domains oriented radially, and the inner core - one large domain, uniformly magnetized along the wire, are visible. This result reproduces what is known about the domain structure of wires with positive magnetostriction $\lambda$ \cite{kino,prb,oli}.

\begin{figure}[h]%
\centering
\includegraphics[width=.49\textwidth]{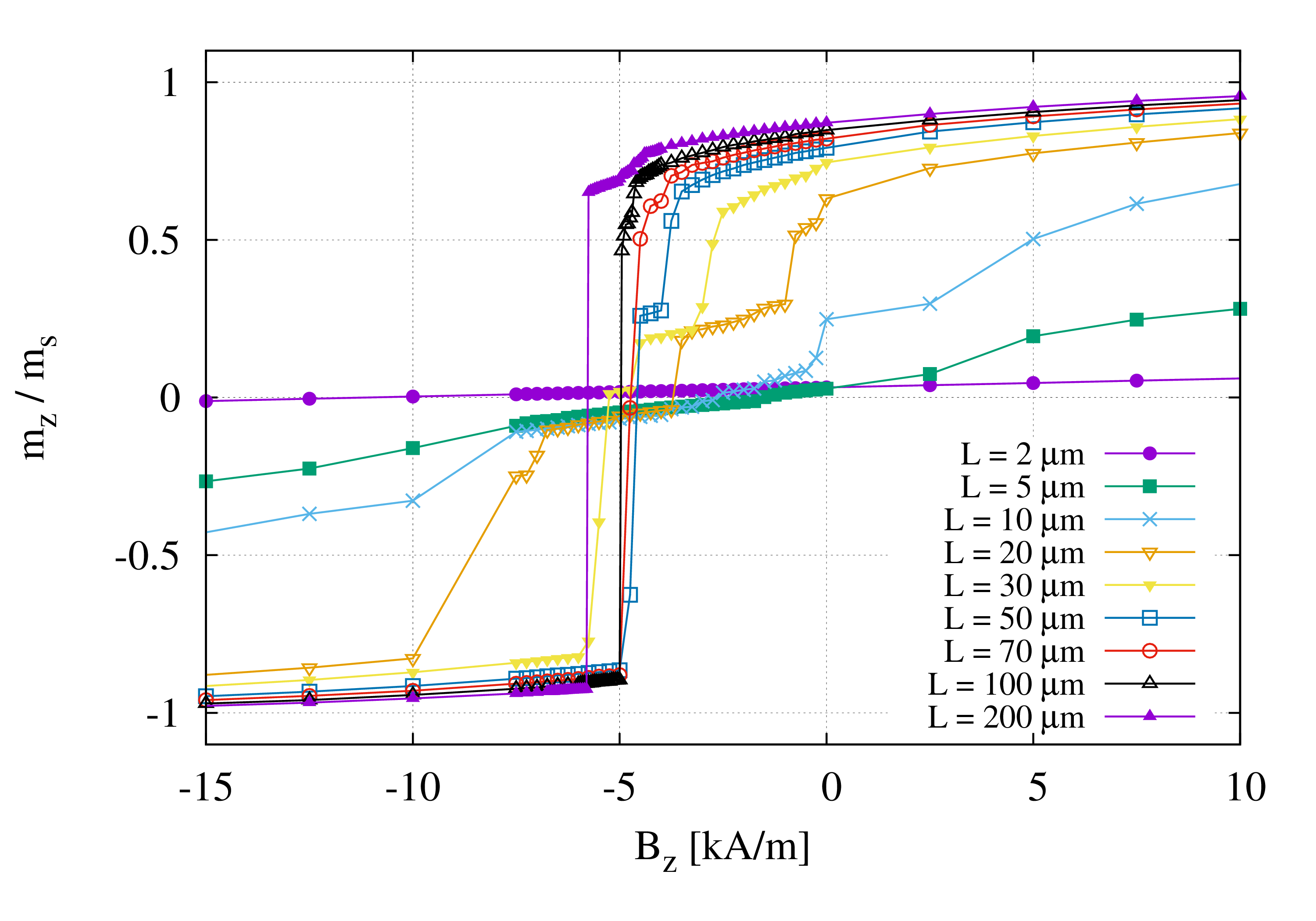}
\caption{The upper branches of the hysteresis loops for a straight wire, for different wire lengths. }
\label{hlp_length}
\end{figure}

\begin{figure}[h]%
\centering
\includegraphics[width=.49\textwidth]{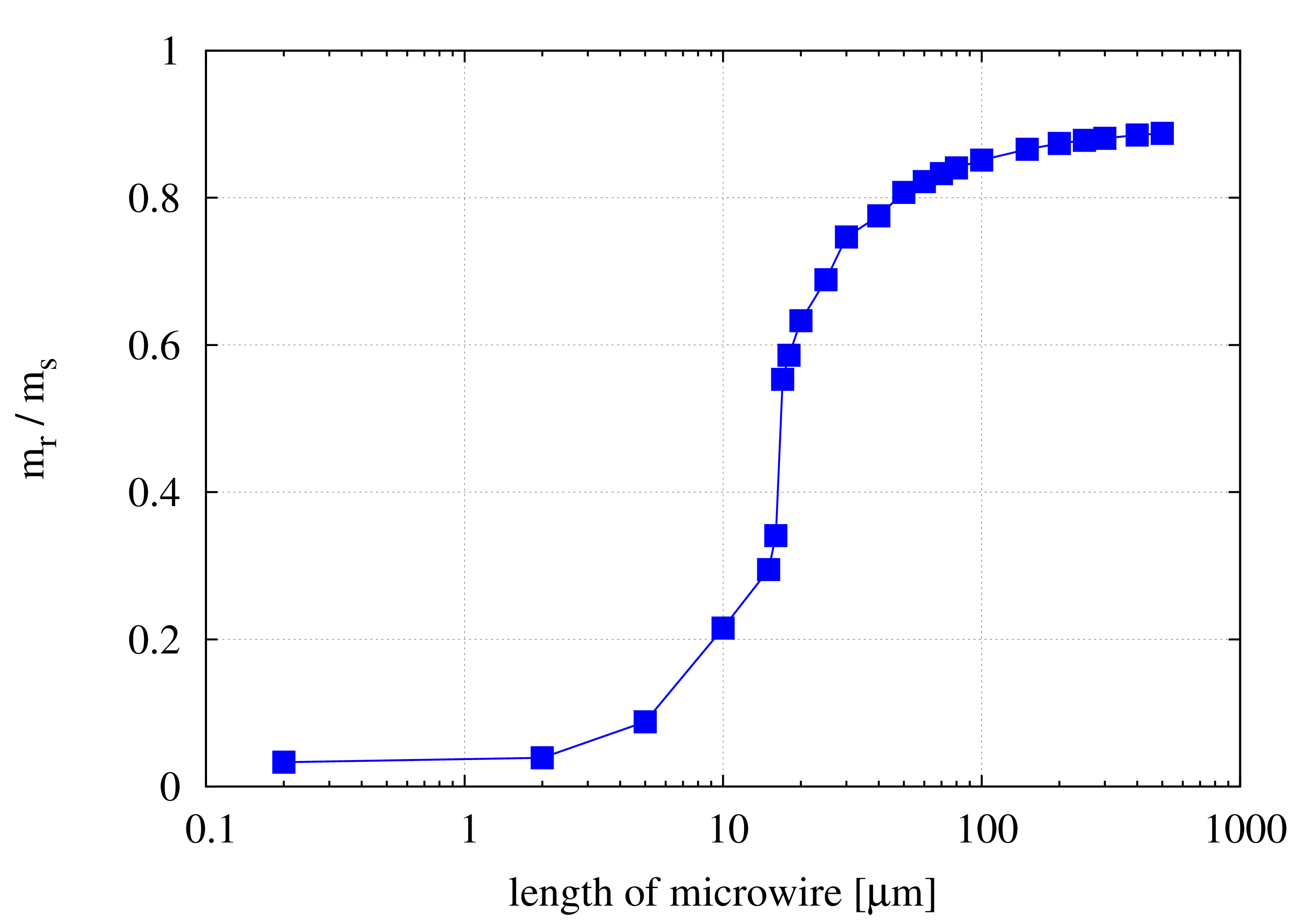}
\caption{The remanence magnetization for a straight wire, as dependent on the wire length - the latter in logarithmic scale. The transition is visible from non-bistable to bistable behavior, when the wire length exceeds some critical value.}
\label{rem_length}
\end{figure}

\paragraph{} In Fig. \ref{dis2}, the distribution of magnetization is shown for similar cross sections at small but different distances from the wire end. From these pictures, it is clear that the axial component of the magnetization increases with the distance from the wire end. This is consistent with what is known on the closure domains in Fe-rich wires \cite{vv,prb}. Also, we have calculated the hysteresis loops for wires of different lengths. These results are shown in Fig. \ref{hlp_length}, and the remanence as dependent on the wire length - in Fig. \ref{rem_length}. These results show that the bistability emerges for the wire length larger than some critical value, as in experiment \cite{vale}. Yet the values of this critical value depends on the wire diameter, and there is no one-to-one correspondence in the results of the calculations and the experimental data.

\begin{figure}[h]%
\centering
\includegraphics[width=.49\textwidth]{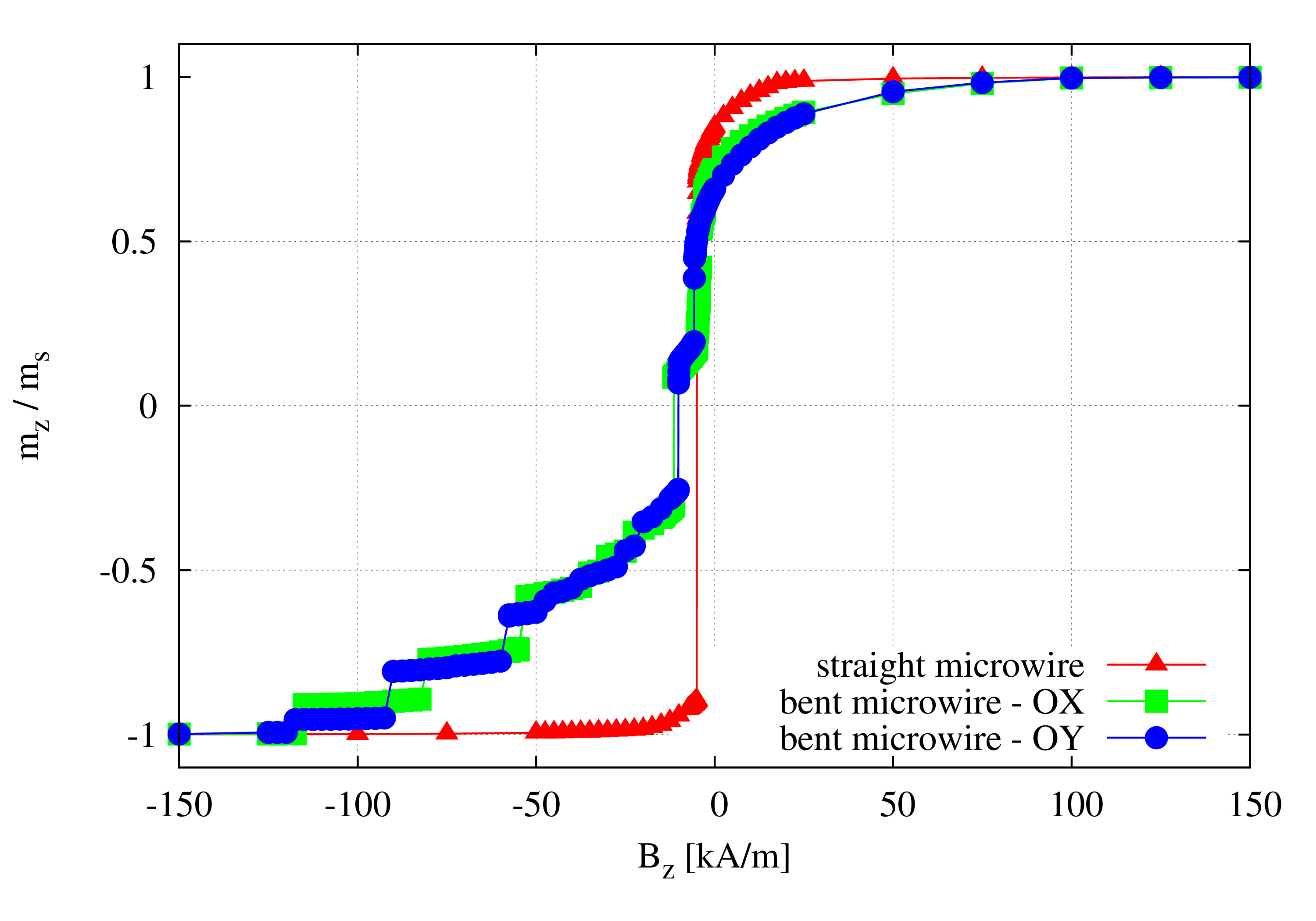}
\caption{The upper branch of the hysteresis loop in a straight wire (red online) and of a bent wire, with two orientations of the easy axis in the compressed part of the wire: OX and OY (green and blue online). The wire radius $r=1 \mu m$, and the length $L=100 \mu m$. For the bent wire, the radius of curvature $R_c= 2 mm$. }
\label{100cc}
\end{figure}

\paragraph{} A test of the approximation of two anisotropy orientations in the compressed volume is to compare the hysteresis loops for both these cases. We show the results of this test in Fig. \ref{100cc}; as a rule, only the upper branch of the loop is shown. In our opinion, the results are satisfactory; two curves approximately coincide and none of them is bistable. On the contrary, the bistability is present for the straight wire.

\begin{figure}[h]%
\centering
\includegraphics[width=.49\textwidth]{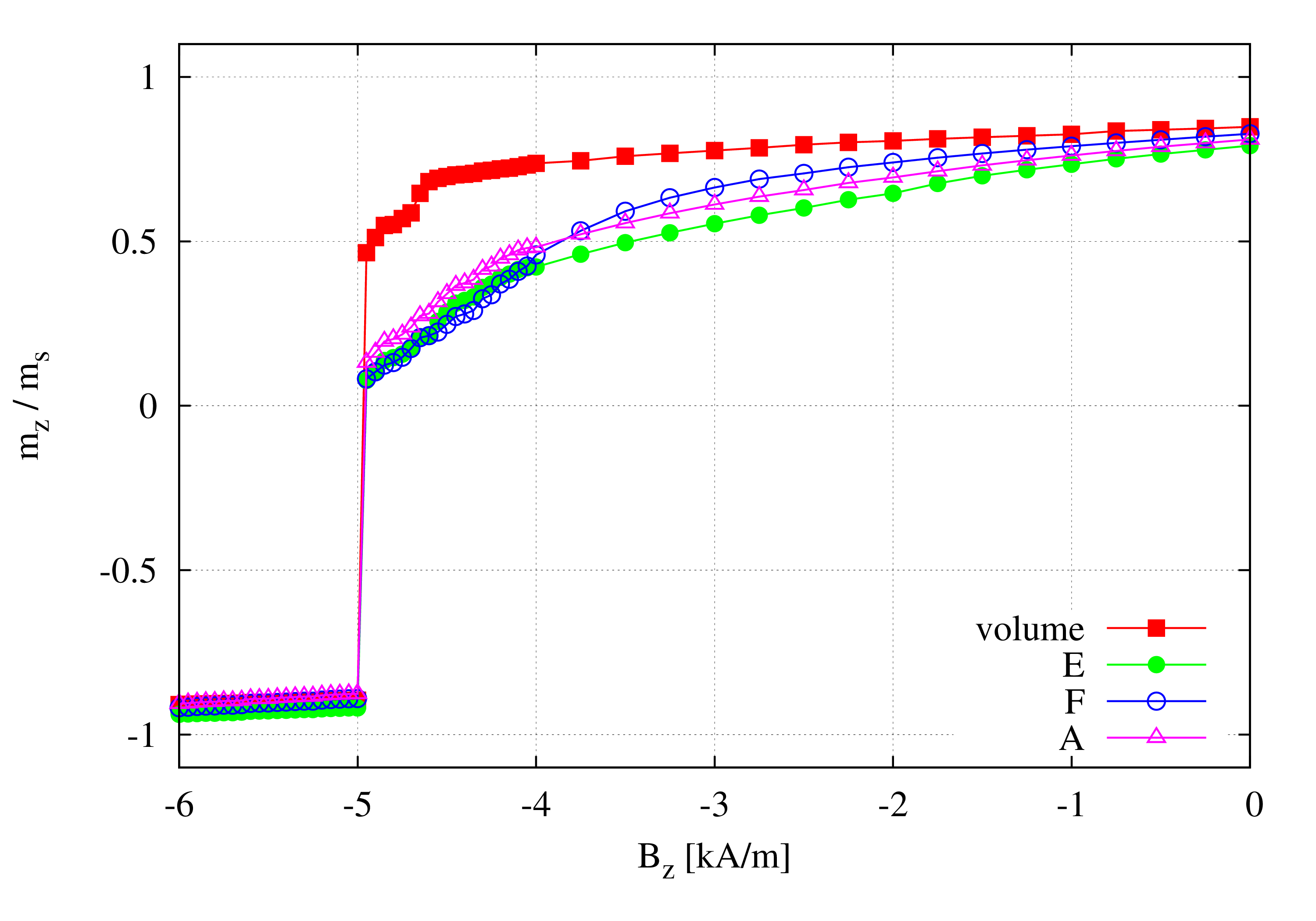}
\caption{The hysteresis loops, volume (red online) and three surface ones (blue, green and violet online) in a straight wire. The wire radius $r=1 \mu m$, the length $L= 100$, and the radius of curvature $R_c=\infty$. }
\label{surfstr}
\end{figure}

\paragraph{} In Fig. \ref{surfstr} we show the volume and surface hysteresis loops for a straight wire. Basically, in the lack of bending stress the three surface loops, taken from three sides, should be the same. Yet, some quantitative differences are visible, which indicate an inaccuracy of the numerical method.

\begin{figure}[h]%
\centering
\includegraphics[width=.49\textwidth]{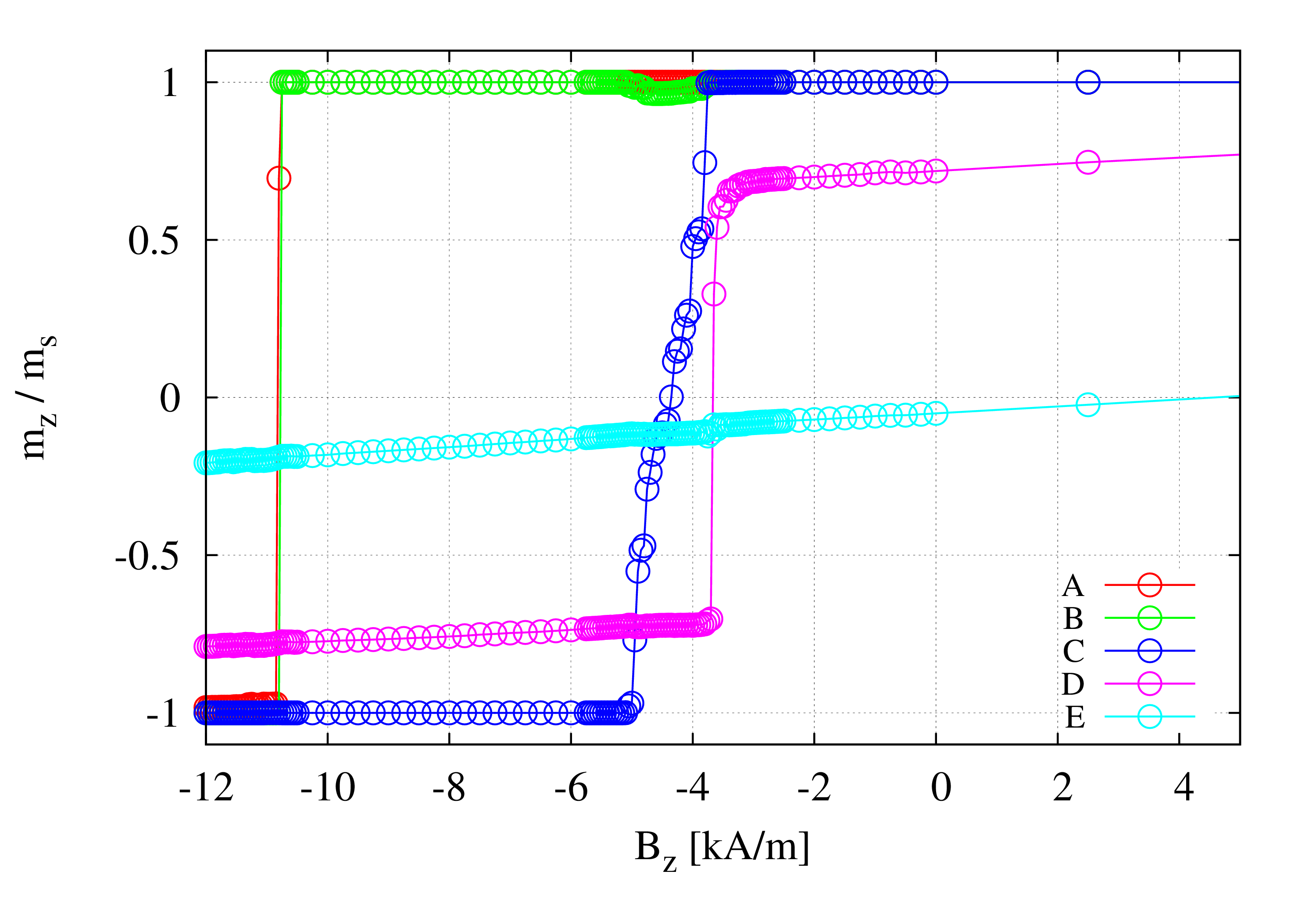}
\caption{The hysteresis loops - contributions from different parts of the wire, from a piece of surface at the concave side (bistable loops, red and green online), to just after the center (dark blue online), to near the surface on the convex, compressed side (violet online) and at the surface (blue online, no hysteresis). The wire radius $r=1 \mu m$, the length $L= 100$, and the radius of curvature $R_c= 2 mm$.}
\label{lefrig}
\end{figure}

\begin{figure}[h]%
\centering
\includegraphics[width=.49\textwidth]{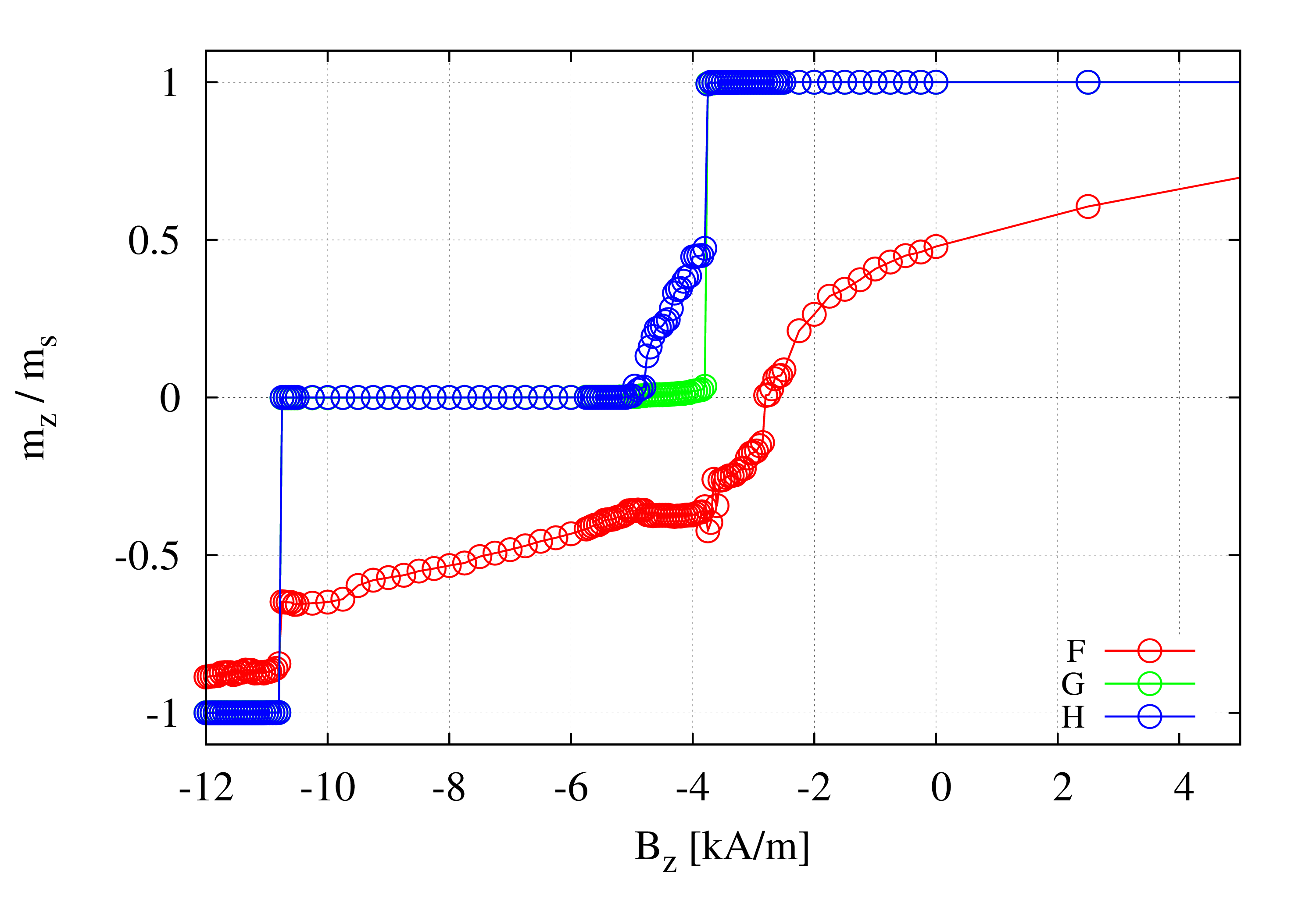}
\caption{The hysteresis loops - contributions from different parts of the wire, from a piece of surface F as shown in Fig. \ref{map2} (red line online), up  to position G (green line online) to the center, position H (dark blue online). The wire radius $r=1 \mu m$, the length $L= 100$, and the radius of curvature $R_c= 2 mm$.}
\label{downup}
\end{figure}

\paragraph{} In Figs. \ref{lefrig} and \ref{downup} we show the hysteresis loops which characterize the magnetic state of some fragments of the wire, on the surface and within the wire volume. The surface loops can be measured by the Kerr method. The internal fragments cannot be measured without removing the surface; however, such removal modifies the stress distribution. Therefore the simulation is the only way of an insight into the magnetic behavior of internal parts of the wire. 

\paragraph{} The results shown in Fig. \ref{lefrig} indicate, that the fragments in the left part of the wire, where the bending stress is tensile, produce bistable contributions to the hysteresis loop. For first two fragments from the left (in the positions A and B in Fig. \ref{map2}), the loops almost coincide, with the coercive field about 11 kA/m. Just on the other side of the wire center, where the bending stress becomes compressive, the coercive field is abruptly reduced (position C in Fig. \ref{map2}, dark blue line online in Fig. \ref{lefrig}). Quite surprisingly, a jump of magnetization is seen as far as in point D; it is only the remanence what is reduced (red line online in Fig. \ref{lefrig}). On the contrary, at the surface (position E in Fig. \ref{map2}, blue line online in Fig. \ref{lefrig})) there is no hysteresis at all; this is because at this area, the local magnetic anisotropy is radial and strong.

\paragraph{} In Fig. \ref{downup}, the contributions to the hysteresis loop are shown for the fragments F, G, and H in the Fig. \ref{map2}. At the position F, the bending stress is neither tensile, nor compressive; its planar contribution is neglected here. The related hysteresis (red line online in Fig. \ref{downup}) is far from bistability. On the contrary, the contribution at point G (within the inner core) shows a 'partial bistability'; apparently, the fragment G consists of two antiparallel domains. Here again we see a difference between the left and right part of the wire. A similar behavior is seen at position H, right in the wire center (dark blue online in Fig. \ref{downup}), with some departure from the preceding case near the coercive field.  


\section{Discussion} In our simulations, two assumptions have been induced by the demands of the software $mumax^3$ and the aim to take into account details of the domain structure in a bent wire. First was the approximation of an effective uniaxial anisotropy, and second - a homogeneous direction of the magnetic anisotropy perpendicular to the wire axis in its compressed part. The quality of the first assumption has been verified by the proper reproduction of the domain structure in the straight wire. To be more precise, this assumption neglects the difference between the actual angular dependence of the energy of the magnetization and the model uniaxial anisotropy where we have just an easy axis. In the latter model, a local magnetic moment can deviate from its easy axis in any direction with the same energy cost, while in real situation a rotation within one plane is easier than within another plane. Yet, the approximation of an effective uniaxial anisotropy does not destroy the domain structure, as seen in Fig. \ref{dis}.

\paragraph{} Our second assumption is that in the compressed part of the wire, the actual direction of the magnetic anisotropy perpendicular to the wire is not relevant. If this is true, we can set this direction along an arbitrary axis without modification of the final results. Actually we have performed this test, as shown in Fig. \ref{100cc}. There, the two curves for the bent wire almost coincide. This means that the assumption does not influence the results qualitatively. Also, our approach is validated by reproduction of two experimental effects: the closure domains and the critical length for the bistability.

\paragraph{} The results of our calculations performed for the bent wire indicate, that below some value of $R_c$, the bistability is destroyed. Here we are not in position to state if the bistability is restored if the radius of curvature is even less, as it was found experimentally in \cite{mv1}. New result of our calculations is that the stress inhomogeneities  inside the bent wire produce a specific domain structure. As for straight wires, this structure is the result of an interplay between the magnetoelastic anisotropy, the exchange and the pseudodipolar interaction. However, in a bent wire the cylindrical symmetry is broken by the bending stress. As a consequence, the inner core is shifted to the convex part, and the compressive bending stress in the concave part of the wire produces a strong anisotropy perpendicular to the wire axis, which destroys the bistability. We conclude that bent wires offer more options of spatial configurations, as ordered and random arrays \cite{gaw,joa}, and therefore 
enable new opportunities of applications. 

\section{Acknowledgement}
The research of P.G. was supported in part by PL--Grid Infrastructure.

%
%

\end{document}